\documentclass[aps,twocolumn]{revtex4}%
\usepackage{amsfonts}%
\usepackage{amsmath}%
\usepackage{amssymb}%
\usepackage{graphicx}
\providecommand{\U}[1]{\protect\rule{.1in}{.1in}}

\newcommand{\bmath}{\begin{displaymath}}
\newcommand{\emath}{\end{displaymath}}
\newcommand{\bite}{\begin{itemize}}
\newcommand{\eite}{\end{itemize}}

\renewcommand{\P}{\mathcal{P}}

\renewcommand{\bell}{{\bf \ell}}

\newcommand{\D}{\mathfrak{D}}

\newcommand{\E}{{\mathcal{E}}}
\newcommand{\Eint}{{\mathcal{E}_{\rm int}}}

\newcommand{\bx}{\mathbf{x}}

\newcommand{\bb}{\mathbf{B}}

\renewcommand{\O}{\mathcal{O}}

\newcommand{\bel}[1]{\begin{equation}\label{#1}}
\newcommand{\bal}[1]{\begin{eqnarray}\label{#1}}
\newcommand{\ee}{\end{equation}}
\newcommand{\ea}{\end{eqnarray}}
\newcommand{\equ}[1]{~Eq.(\ref{#1})}

\newcommand{\Tr}{{\rm Tr}}

\usepackage{graphicx}
\usepackage{latexsym}
\usepackage{amssymb}

\begin{document}
\preprint{HEP-TH/1000708}
\title[Repulsive Casimir]{On the Direction of Casimir Forces}
\author{Martin Schaden}
\affiliation{Department of Physics, Rutgers University, 101 Warren Street, Newark NJ 07102}

\begin{abstract}

The Casimir force due to a massless scalar field satisfying Dirichlet
boundary conditions may attract or repel a piston in the neck of a flask-like container.
Using the world-line formalism this behavior is related to the competing contribution to the interaction 
energy of two types of Brownian bridges. It qualitatively is also expected from attractive long-range
two-body forces between constituents of the boundary. A geometric subtraction scheme is presented that avoids 
divergent contributions to the interaction energy and classifies the 
Brownian bridges that contribute to the force. These are all of finite length and the Casimir
force can be analyzed and in principle accurately computed without resorting to
regularization or analytic continuation. The world-line analysis is robust with respect to
variations in the shape of the piston and the flask and the analogy with long-range forces 
suggests that neutral atoms and particles are also drawn into open-ended pipes (or nano-tubes) 
by Casimir forces of electromagnetic origin.

\end{abstract}





\startpage{101}
\endpage{120}
\maketitle

\section{Introduction}

Contrary to intuition derived from the attractive Casimir force between two conducting
plates\cite{Casimir48}, Boyer\cite{Boyer68} found that the zero-point energy apparently tends
to expand a perfectly conducting spherical shell. Until recently\cite{Schaden06}, there was no
qualitative explanation for the sign of the Casimir energy. However, the finite negative surface
tension of a metallic spherical shell cannot be measured by itself: changing the radius
of a real cavity necessarily involves its material properties. The negative Casimir tension of a
spherical shell in this sense is a result without direct physical implications. It was later
found that Casimir self-energies of many closed cavities are plagued by divergences that
cannot be removed without appealing to material properties of their walls\cite{Deutsch79}.

The Casimir force between disjoint solid bodies on the other hand in principle is observable and
ought to be finite. For some simple shapes the force between uncharged conductors has
now been measured quite accurately \cite{Experiments}. Experimentally as well as theoretically the
force between conductors is attractive in all cases studied. A theorem by Kenneth and Klich
\cite{Klich06} and its generalization by Bachas \cite{Bachas06} states that reflection positivity
implies that the interaction between a mirror-pair of disjoint (charge-conjugate) bodies is
attractive. This theorem in particular implies that, contrary to previous
suggestions\cite{Elizalde91,Lamoreaux97}, the Casimir force between two half-spheres is
attractive\cite{Klich06}. The attractive Casimir-Polder\cite{CasimirPolder48} force between
polarizable atoms furthermore suggests that the force could be attractive for \emph{any} shape of
the conductors. Such considerations, as well as many failed attempts
\cite{Elizalde91,Lamoreaux97,Schmidt06} to find shapes exhibiting  repulsion might give the impression that
repulsive Casimir forces between distinct bodies occur for suitable (mixed) boundary
conditions\cite{Hushwater97,Barton06,Fulling07} only.

But neither the long list of examples nor the restrictive theorems by Kenneth, Klich and
Bachas\cite{Klich06,Bachas06} apparently imply that the Casimir force is attractive between
\emph{any} conductors. Intuition based on the Casimir-Polder\cite{CasimirPolder48} force between
atoms could be misleading\cite{Kenneth02}. Polarizable atoms attract just as any distant conducting
spheres would and as such do not even qualitatively reproduce the Casimir energy of some
geometries. [If vacuum forces are entirely due to attractive two-body forces, a metallic spherical
shell apparently would have to have positive surface tension.]

Semiclassical\cite{Mateescu07} and numerical\cite{Schaden08} arguments suggest that the Casimir
force on a piston depends qualitatively on the shape of the casing. We will see below that a piston
in the neck of a flask with a spherical bulb in fact may be attracted or repelled from the
bulb and can have a stable equilibrium position. This will be shown for the Casimir force on the
piston due to a massless scalar field satisfying Dirichlet boundary conditions on the flask and
piston surfaces. Although the world-line approach we use is for a scalar field, the results
can be qualitatively understood as due to an attractive long-range interaction between
constituents of the boundaries. The outward Casimir pressure on an ideal metallic sphere\cite{Boyer68,Milton78} and
inward pressure on an ideal metallic cylinder\cite{DRM81} also suggest that a similar competition
of vacuum forces occurs in the electromagnetic case with metallic boundaries. Net Casimir forces
that change direction or vanish perhaps can be observed in micro-mechanical devices. The existence
of stable equilibrium positions with a vanishing Casimir force for a large class of shapes could also be of
practical interest in high precision studies of long-range forces.

The method used here to determine the direction of the Casimir force does not require an atomistic
interpretation of its origin, but uses a geometrical subtraction scheme to express finite Casimir
energies as a sum of \emph{finite} contributions of definite sign due to classes of Brownian bridges
of finite length. The need to compute differences of potentially arbitrary large quantities to
obtain Casimir forces is thereby avoided.

\section{World-line approach}
Consider the heat kernel operator $\mathfrak{K}_\D(\beta)=e^{{\beta\triangle}/2}$ for the Laplacian
$\triangle$ with Dirichlet boundary conditions on a bounded domain $\D\subset\mathbb{R}^3$. The
eigenvalues $\{\lambda_n\ge\lambda_{n-1}>0;n\in\mathbb{N}\}$ of the negative Laplace operator in
this case are discrete, real and positive and the corresponding spectral function (or trace of the
heat-kernel),
\bel{spfunc}
\phi_\D(\beta)=\Tr\mathfrak{K}_\D(\beta)= \sum_{n\in\mathbb{N}} e^{-\beta\lambda_n/2},
\ee
is finite and well defined for $\beta>0$. In principle, the spectral function includes all the
information required to compute zero point energies of bounded domains.

$\phi_\D(\beta\sim 0)$ has the well-known\cite{Greiner71,Gilkey84,FK88} asymptotic (short time or
high-tem\-per\-a\-ture) expansion,
\bel{assfunc}
\phi_\D(\beta\sim 0)\sim \frac{1}{(2\pi\beta)^{3/2}}\sum_{n=0}^\infty (2\pi\beta)^{n/2} a_n(\D)
+\O(e^{-l^2/\beta}).
\ee
For smoothly bounded domains, the Hadamard-Minakshisundaram-DeWitt-Seeley coefficients $a_n(\D)$ in
this series are integrals over powers of the local curvature and reflect average geometric
properties of the domain and its boundary\cite{kac66,Vassilevich02}. For a bounded
three-dimensional flat Euclidean domain $\D$, $a_0(\D)$ gives its volume $\mathcal{V_\D}$ and
$a_1(\D)=-\mathcal{S_\D}/4$ gives the surface area $\mathcal{S_\D}$ of its boundary\cite{kac66}.
$a_2(\D)$ is proportional to the integrated curvature [sharp edges of the boundary also contribute
\cite{SW71, Balian}] and $a_3(\D)$ is a dimensionless coefficient reflecting topological characteristics of the domain [such as the connectivity of its boundary and the number and opening
angles of its corners\cite{kac66, Balian}]. The coefficient $a_4$ is the most crucial
for Casimir effects, since $a_4(\D)\neq 0$ implies a logarithmic divergent vacuum energy that
prevents one from uniquely defining the Casimir energy. The geometric origin of this
coefficient\cite{Vassilevich02} is, however, not simple to describe. Non-analytic and
(for $\beta\sim 0$) exponentially suppressed contributions to the asymptotic expansion of
$\phi_\D(\beta)$ are associated with classical periodic- and diffractive- orbits\cite{Brackbook} of
a minimal length $l$.

The world-line approach to Casimir energies\cite{GLM03} is based on the observation\cite{kac66,
Stroock93} that the spectral function for a bounded flat Euclidean domain $\D$ can be expressed in
terms of its support of standard Brownian bridges. In three dimensions,
\bel{support}
\phi_\D(\beta)=\int_\D \frac{d\bx}{(2\pi \beta)^{3/2}} \P[\bell_\beta(\bx)\subset\D ]\ ,
\ee
where $\bell_\beta(\bx)=\{\bb_\tau(\bx,\beta), 0\leq \tau\leq \beta;
\bb_0(\bx,\beta)=\bb_\beta(\bx,\beta)=\bx\}$ is a standard Brownian bridge from $\bx$ to $\bx$ in
"proper time" $\beta$ and $\P[\bell_\beta(\bx)\subset\D ]$ denotes the probability for the bridge
to be entirely within the bounded domain $\D$.

Although the spectral function of a bounded domain of finite volume thus is evidently finite, divergences arise 
in the corresponding zero-point energy. The formal zero-point energy of a massless scalar
satisfying Dirichlet boundary conditions on $\D$,
\bel{Ecas}
\E_{\rm vac}(\D)\sim{\frac 1 2}\sum_{n=0}^\infty\sqrt{\lambda_n}\sim-\pi\int_0^\infty
\frac{d\beta}{(2\pi\beta)^{3/2}}\phi_\D(\beta)\ ,
\ee
diverges due to the behavior of the integrand for $\beta\sim 0$ implied by \equ{assfunc}. To calculate
finite Casimir energies one customarily regulates the integral in \equ{Ecas} [for instance by
analytic continuation or with a finite lower limit in the integration over
the proper time in\equ{Ecas}]. One then must show that the physical effect of interest
remains finite after analytic continuation or when the cutoff is removed.
The same result can also be achieved using a numerically more suitable
and physically more transparent cutoff-independent procedure by noting that only the first
few terms (first five in three spatial dimensions) of the asymptotic high-temperature expansion
in\equ{assfunc} lead to divergent contributions to the vacuum energy. The behavior of the integrand for
$\beta\rightarrow 0$ therefore can be improved by considering a (finite) linear combination of spectral
functions for domains $D_k, k=0,1\dots$,
\bel{tildephi}
\tilde\phi(\beta)=\sum_{k} c_k \phi_{\D_k}(\beta)\ .
\ee
When the coefficients $c_k$ are chosen so that
\bel{finiteness}
\sum_{k} c_k a_n(\D_k)=0\ \ {\rm for}\ \ n=0,\dots,4 \ ,
\ee
the "interaction" vacuum energy,
\bel{inter}
\Eint=-\pi\int_0^\infty \frac{d\beta}{(2\pi\beta)^{3/2}}\tilde\phi(\beta)=\sum_{k} c_k \E_{\rm
vac}(\D_k)\ ,
\ee
is finite because the integrand of\equ{inter} is $\O(\beta^{-1/2})$.  Due to the geometrical nature
of the asymptotic heat kernel expansion, the linear combination $\Eint$ of zero-point energies
$\E_{\rm vac}(\D_k)$ may be interpreted as the difference in vacuum energy for
domains with the same total volume, total surface area, average curvature, topology, etc\dots  It
of course is here understood that the subtractions at most affect the physical quantity of interest in
a calculable way. The most convenient subtractions thus could depend on the physical problem at hand.

In the context of Casimir effects, such a "geometric" scheme was first used by
Power\cite{Power64} to derive the original Casimir force\cite{Casimir48} between parallel metallic
plates without intermediate regularization. Power compared the vacuum energy of a
metallic box of fixed dimensions with a moveable plate for different positions of the plate.
Svaiter \cite{Svaiter92} recognized and succinctly emphasized the physical nature of this scheme.
In\cite{Schaden06} finite contribution to the interaction vacuum energy due to periodic orbits were
computed in leading semiclassical approximation. However, diffractive orbits and non-vanishing
higher terms in the asymptotic power series $\tilde\phi(\beta)$ may sometimes contribute significantly to
$\Eint$ \cite{Schaden08}.
\newcounter{figures}
\begin{figure}
\refstepcounter{figures}\label{fig1}
\includegraphics[width=3.4in]{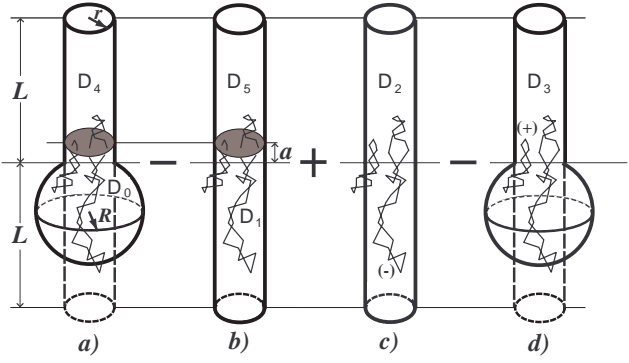}
\label{pistons}
\caption{\small The interaction Casimir energy $\Eint(a)$ for a piston
in the cylindrical neck [of radius $r\le R$ and length $L>R$] for a flask  with a spherical bulb [of radius $R$].
The force on the piston in the flask at height $a$ above its bulb is compared to that on a piston displaced a distance $a$ from the center of a cylinder of overall length $2L$ and the same radius $r$. The relevant interaction energy $\Eint(a)$ is the alternating sum of
zero-point energies of a massless scalar field satisfying Dirichlet's condition on the boundaries of domains $\D_0$ to $\D_5$. The cylindrical extension
of the neck indicated by dashed lines in a) and c) should guide the eye and is {\it not} part of the flask. the (+)- and (--)-loops shown are examples of Brownian bridges that contribute positively/negatively to $\Eint(a)$: (+)-loops contribute only to the spectral
function $\phi_{\D_3}(\beta)$, whereas (-)-loops contribute only to $\phi_{\D_2}(\beta)$.}
\end{figure}

\section{The Casimir force on a piston in the neck of a flask}
The world-line formalism of the previous section can be used to obtain the Casimir force on the
piston in the flask of Fig.~1a) with a bulb of radius $R$,  and a neck of radius $r<R$ and length
$L>R$.  To avoid the dangerous first five terms in the asymptotic expansion of $\tilde\phi(\beta)$
consider the linear combination
of spectral functions for the bounded domains shown in Figs.~1a)-1d). The vacuum energies of
Fig.~1c)~and~1d) do not depend on the height $a$ of the piston, but those of Fig.1a) and 1b) do.
For any finite $L>R$, the interaction energy,
\bal{Eflask1}
\Eint&=&\sum_{k=0}^3(-1)^k\E_{\rm vac}(\D_k)\nonumber\\
&=&-\frac{1}{8\pi^2}\int_0^\infty\frac{d\beta}{\beta^3}\int d\bx \sum_{k=0}^3 (-1)^k
\P[\bell_\beta(\bx)\subset\D_k ]\ ,\nonumber\\
&&\hspace{-4em}\D_0={\rm flask~below~piston},\ \D_1={\rm cylinder~below~piston},\nonumber\\
&&\hspace{-4em}\D_2={\rm whole~cylinder},\ \D_3={\rm whole~flask},\nonumber\\
&&\hspace{-4em}\D_4=\D_5={\rm cylinder~above~piston}
\ea
only allows one to compute the \emph{difference} in the force on the piston within the
neck of the flask and within a cylinder of length $2 L$ of the same radius $r$. For finite $L>R\ge
r>0$ the domains $\D_0\dots\D_5$ are all bounded, but we will be especially interested in the limit
of large $L/R$  where the force on the piston near the center of the cylinder is negligible.
Eqs.(\ref{support}),~(\ref{tildephi})~and~(\ref{inter}) give the second expression of
$\Eint$ in\equ{Eflask1}.  The volumes $\D_4$ and $\D_5$ above the piston in the flask and in the
cylinder are identical and give no net contribution to the alternating sum in \equ{Eflask1}. By conditioning on whether a loop pierces certain surfaces one can show that the so defined interaction vacuum energy is finite for any height $0<a<L$ of the piston. Only bridges that
pierce (or touch) the piston contribute to the alternating sum in\equ{Eflask1}: if they are
entirely within $\D_0$ or $\D_4$ (the parts of the flask below and above the piston) then they also
are entirely within $\D_3$ (the whole flask) and if they are entirely within the cylinder below or
above the piston ($\D_1$ or $D_5$) then they also are within the whole cylinder ($\D_2$). A loop
that \emph{only} pierces the piston, or in addition pierces \emph{both} cylinder and flask also
gives no contribution. Only two types of Brownian bridges (shown schematically in Fig.1) therefore contribute
to $\Eint$. They either
\begin{description}
          \item[(+)] pierce piston and cylinder, but not the flask,
          \item[or]\hfill (I)
          \item[(--)] pierce piston and flask, but not the cylinder.
\end{description}
The shortest bridge that contributes to $\Eint$ is of type~(+) and has extent  $a$: it just
touches the piston and pierces the cylinder (but not the flask) near the base of the neck. The
length of all loops that contribute to $\Eint(a>0)$ thus is bounded below by $2 a$.
For $\beta\rightarrow 0$ the probability of a loop of finite extent is exponentially small. The
asymptotic expansion of $\tilde\phi(\beta)$ for $\beta\sim 0$ therefore vanishes to all orders
and $\Eint$ is finite.

A slight elaboration on the previous argument gives the direction of the force on the piston
for some extreme configurations. Bridges of type~(+) are within the whole flask (domain $\D_3$) but not entirely
within any of the other five domains. They therefore give a \emph{positive} contribution to
$\Eint$. Bridges of type~(--) on the other hand are within domain $\D_2$ (the whole cylinder) only
and give a \emph{negative} contribution to $\Eint$ in\equ{Eflask1}. The sign of $\Eint$ thus
depends on which of the two types of bridges occurs more frequently and can be readily established 
for the following cases.
\medskip

\noindent $L>R=r$
\newline Since any loop that pierces the cylinder also pierces this flask with a hemispherical bottom, there is no contribution from (+)~loops and $\Eint$ is negative \cite{Schaden08} for any height $a>0$.
\medskip

\noindent $a\ll r\ll R\ll L/2$:
\newline For this flask with a very long, thin neck, 
the probability of a bridge over time $\beta$ of type~(--) is much less than one of type~(+) when the piston is near the base of the neck. (--)-loops in this case have a length greater than $2 R+a$ but a
transverse extent of less than $2 r$ and are very elongated. Their contribution, $\Eint^{(-)}$, to $\Eint$
may be estimated by noting that longitudinal and
transverse components of a bridge are statistically independent. The probability that a bridge
starting and ending at $\bx=(\bx^\bot,z)$ lies entirely within a cylinder, $C(r,l)$, of radius $r$
and length $l$ therefore is the product of the probabilities for the transverse bridge to remain within the disk
$D(r)$ of radius $r$, and for the one-dimensional longitudinal bridge to lie within the interval
$[0,l]$,
\bel{decompose}
\P[\bell_\beta(\bx)\subset C(r,l)]=\P[\bell_\beta^\bot(\bx^\bot)\subset D(r)]
\P[\ell_\beta^\|(z)\subset [0,l]]\ .
\ee
Ignoring (for $R\gg r$ small) corrections due to the curvature of the flask
bottom, the contribution $\Eint^{\hspace{-1em}(-)}$ of bridges of type~(--) in this regime becomes,
\bal{separate}
\Eint^{\hspace{-1em}(-)}(a;L\gg R\gg r)&\sim& -\int_0^\infty \hspace{-1em}d\beta \phi_{D(r)}(\beta)\nonumber\\
&&\hspace{-4em}\times\int^\infty_{2 R+a}\hspace{-1em}ds\frac{(s\!-\!2 R\!-\!a)}{4\pi\beta^2} \frac{d\P[|\ell_\beta^\||<s]}{ds} \nonumber\\
&&\hspace{-10em}=-\int_0^\infty\hspace{-1em}d\beta\phi_{D(r)}(\beta)\int^\infty_{2 R+a}
\hspace{-1em}ds \frac{(s\!-\!2 R\!-\!a)}{2\sqrt{2\pi}\beta^{3/2}}
\frac{d^2\phi_{[0,s]}(\beta)}{ds^2}
\nonumber\\
\ea
Here $|\ell_\beta |$ denotes the maximal extent of a standard (one-dimensional) Brownian bridge
over time $\beta$ [the overall extension of a bridge evidently does not depend on its starting
point]. The factor $(s-2R-a)$ in\equ{separate} is the translational volume available to a
one-dimensional loop of extent $s$ that pierces the piston (at height $a$) as well as the bulb at
the bottom of the flask (at height $-2R$). $\phi_{D(r)}=\int d\bx \P[\bell_\beta^\bot(\bx)\subset
D(r)]/(2\pi\beta)$ is the spectral function for the disk $D(r)$. The last line of \equ{separate}
uses that the spectral function of the interval $[0,s]$ is $\phi_{[0,s]}(\beta)=\int_0^s dx
\P[|\ell_\beta|<x]/\sqrt{2\pi\beta}$. The spectral functions of an interval and of a disk are
known\cite{Brackbook} but for our estimate it suffices that the analog of \equ{support}
for two dimensions\cite{kac66} implies that $\phi_{D(r)}(\beta)<r^2/(2\beta)$ and that
\bel{interval}
\phi_{[0,s]}(\beta)=\frac{s}{\sqrt{2\pi\beta}}\left(1+2\sum_{n=1}^\infty e^{-2 s^2
n^2/\beta}\right)\ .
\ee
$\Eint^{\hspace{-1em}(-)}$ for $R\gg r$ thus is bounded by,
\bal{bounds}
0&>&\Eint^{\hspace{-1em}(-)}(a;L\gg R\gg
r)>\nonumber\\&&\hspace{-2.5em}-\hspace{-.5em}\sum_{n=1}^\infty\int_0^\infty\hspace{-1em}d\beta\frac{
r^2(2 R+a)e^{-\frac{2}{\beta} (n(2 R+a))^2}}{4\pi\beta^3} =\frac{-\pi^3 r^2}{1440(2R+a)^3}\
.\nonumber\\&&
\ea
The lower bound in\equ{bounds} is just the Casimir energy due to a massless scalar associated with
two parallel flat plates of area $\pi r^2$ that are separated by a distance $d=2R+a$. [This lower
bound of $\Eint^{\hspace{-1em}(-)}$ remarkably holds for $d\gg r$!]

The sign of $\Eint$ for finite $a/r$ in the limit $r/R\rightarrow 0$ follows from the fact that $\Eint^{\hspace{-1em}(-)}$ vanishes in
this limit whereas the corresponding positive contribution,
$\Eint^{\hspace{-1em}(+)}$, of (+)-loops does not (it even slightly increases for $R\rightarrow
\infty$. For any piston height, $a$, and neck radius, $r$, the interaction energy $\Eint$ therefore is
positive when $L$ and $R$ are sufficiently large,
\bal{limits}
\Eint^{\hspace{-1em}\infty}(a;r)&=&\lim_{R\rightarrow \infty}\lim_{L\rightarrow
\infty}(\Eint^{\hspace{-1em}(+)}+\Eint^{\hspace{-1em}(-)})\nonumber\\ &\rightarrow &
\Eint^{\hspace{-1em}(+)}(a,r;L>R\rightarrow\infty)>0\ .
\ea

\noindent $L> a \gg r $:
\newline The piston is very high in the neck of the flask. The magnitudes of $\Eint^{\hspace{-1em}(-)}(a,r)$ and
$\Eint^{\hspace{-1em}(+)}(a,r)$ \emph{both} vanish at least as fast as $(r/a)^{3}$.
Since \emph{both} types of loops are highly elongated in the limit of large $a/r$,  this can be seen
by slightly adapting the previous proof that the $\Eint^{\hspace{-1em}(-)}$ contribution vanishes
for large values of $r/(2R+a)$, . We therefore have that
\bel{limits1}
\Eint(L>a\gg r)\sim 0\ .
\ee
\equ{limits1} together with \equ{limits} implies that $\Eint^{\hspace{-1em}\infty}(a;r)$
\emph{cannot} be monotonically increasing with piston height $a$. For sufficiently large $L$ and $R$, the force on the piston at some positions (close to the bulb), is directed
away from the (large) bulb. The piston in this this sense is repulsed by the bulb, but it probably is more 
accurate to say that the piston is being drawn into the thin neck of such a flask.
Whichever point of view is adopted, the upward force on the piston of Fig.~1 extends to all
$a$ in the limit $L>R\rightarrow\infty$ (i.e. for a half-space with a hole of radius $r$), because
the contribution $\Eint^{\hspace{-1em} -}$ is negligible and $\Eint^{\hspace{-1em} +}$
evidently decreases monotonically with the piston height $a$. This repulsion thus is \emph{not} directly
related to the negative vacuum contribution to the surface tension of a spherical shell. The latter 
vanishes inversely proportional to $R^2$ (for metallic-\cite{Boyer68} and
semiclassically\cite{Schaden06b} also for Dirichlet- boundary conditions). It rather is the 
net vacuum force on the piston due to the flask bulb and the long thin cylindrical neck.
The latter is finite even if the force due to the spherical bulb may be neglected.

\section{Conclusion}
The direction of the Casimir force on a body can depend relatively sensitively on the shape of the
surrounding boundary. Such shape dependence was previously conjectured\cite{Schmidt06} on the basis of a change in sign of the Casimir self-energy of a parallelepiped\cite{Lukosz71}. However, it was shown\cite{Jaffe05} and has been proven\cite{Klich06,Bachas06} on more
general grounds that the Casimir \emph{force} always attracts the piston to the nearest 
end-plate in this case.
We here examined more asymmetric piston configurations for which reflection 
positivity\cite{Klich06,Bachas06} does not imply the direction of the force. 
We used the world-line formalism and a geometrical subtraction scheme to determine the Casimir force on the piston due to a massless scalar field satisfying Dirichlet boundary conditions in these more complicated geometries. Although numerical calculations in principle are possible in this scheme, the quantitative results depend strongly on the precise geometry and are of limited practical interest. We therefore restrict ourselves to a qualitative analysis that applies to a large class of geometries. Our considerations in fact do not depend on the details of the shape of the "flask" and of the "piston" and can be extended to pistons that are not disks and/or do not touch the cylindrical neck. The bulb of the flask  also may be replaced by a more general cavity with average dimensions that are larger than the neck radius. The force on a piston in the neck of such a "flask" still depends on competing contributions to the Casimir energy
from just two types of Brownian bridges (with properties given in (I)). Since these are of finite extent, their contributions to the interaction energy can be estimated and compared. For the flask of Fig.~1, we find that at any given height $a$ the direction of the Casimir force on the piston depends on the ratio $r/R$ of the neck's radius to that of the bulb (assuming the neck is sufficiently long for its end to be ignored). The piston is always drawn into the neck for very small values of $r/R$, whereas it is always attracted to the bulb of a flask with hemispherical bottom $r/R=1$. We conclude that for finite values of the ratio $0<r/R<1$, the Casimir force on the piston vanishes at some (finite) height $a$. The existence of such an equilibrium position in some geometries perhaps is of interest to high precision measurements of forces.  These results based on the world-line formalism can be qualitatively understood as due to long-range \emph{attractive} two-body forces between constituents of the boundary. Since Van-der-Waals and Casimir-Polder\cite{CasimirPolder48} forces between neutral atoms and molecules are of this nature, we expect them to qualitatively also obtain in realistic systems.

\noindent{\bf Acknowledgements:}This work was supported by the National Science Foundation with
Grant~PHY0555580.

\end{document}